\documentclass[12pt,preprint]{aastex}

\def\rmmat#1{{\hbox{\rm #1}}}
\def\rmscr#1{\rmmat{\scriptsize #1}}
\newcommand{\be}{\begin{equation}}
\newcommand{\ee}{\end{equation}}
\newcommand{\bt}{\begin{table} \begin{center}}
\newcommand{\et}{\end{center} \end{table}}
\newcommand{\ba}{\begin{eqnarray}}
\newcommand{\ea}{\end{eqnarray}}
\newcommand{\ie}{{\it i.e.~}}
\newcommand{\eg}{{\it e.g.~}}

\def\eqref#1{Equation~(\ref{eq:#1})}

\def\figref#1{Figure~\ref{fig:#1}}

\begin{document}
\newcommand{\bfi}{{\bf B}} \newcommand{\efi}{{\bf E}}
\newcommand{\lel}{{\lambda_e^{\!\!\!\!-}}}
\newcommand{\me}{m_e}
\newcommand{\mcs}{{m_e c^2}}
\def\ho{{\hat {\bf o}}}
\def\hm{{\hat {\bf m}}}
\def\hx{{\hat {\bf x}}}
\def\hy{{\hat {\bf y}}}
\def\hz{{\hat {\bf z}}}
\def\hom{{\hat{\mathbf{\omega}}}}
\def\hr{{\hat {\bf r}}}
\def\omv{\mathbf{\omega}}

\title{R-Modes on Rapidly Rotating, Relativistic Stars: I. Do
Type-I Bursts Excite Modes in the Neutron-Star Ocean?}

\author{Jeremy S. Heyl\altaffilmark{1}}
\affil{
Harvard-Smithsonian Center for Astrophysics, MS-51, 
60 Garden Street, Cambridge, Massachusetts 02138, United States}
\email{heyl@physics.ubc.ca}
\altaffiltext{1}{Chandra Fellow, Current Address: Department of Physics and Astronomy,
6224 Agricultural Road, Vancouver BC V6T 1Z1, Canada}

\begin{abstract}
During a Type-I burst, the turbulent deflagation front may excite
waves in the neutron star ocean and upper atmosphere with frequencies,
$\omega \sim 1$~Hz.  These waves may be observed as highly coherent flux
oscillations during the burst.  The frequencies of these waves changes
as the upper layers of the neutron star cool which accounts for the
small variation in the observed QPO frequencies.  In principle several
modes could be excited but the fundamental buoyant $r-$mode exhibits
significantly larger variability for a given excitation than all of
the other modes.  An analysis of modes in the burning layers themselves 
and the underlying ocean shows that it is unlikely these modes can account 
for the observed burst oscillations.  On the other hand, photospheric modes 
which reside in a cooler portion of the neutron star atmosphere may provide an
excellent explanation for the observed oscillations.
\end{abstract}

\keywords{stars : neutron, oscillations -- X-rays : bursts, binaries}

  \section{Introduction}

  As material accretes onto the surface of a star, the generation of
  nuclear energy may be stable or unstable depending on the rate of
  accetion and the properties of the underlying star.  Unstable nuclear
  burning on the surface of a neutron star manifests itself as Type-I
  X-ray bursts
  \citep[][the final two are reviews]{1975ApJ...195..735H,1976ApJ...205L.127G,1976Natur.263..101W,1977Natur.270..310J,1978ApJ...220..291L,1993SSRv...62..223L,Stro03}.
  The accumulation of material on the surface of the star is likely to
  be sufficiently chaotic that the nuclear burning ignites at a
  particular point, so one would expect that as the burning envelopes
  the stellar surface that the observed flux to vary at approximately
  the spin frequency of the star \citep{Joss78}.
  \citet{2002ApJ...566.1018S} and \citet{2001ApJS..133..195Z} present
  recent models for the growth of the flame other the stellar surface.
  The discovery of these oscillations had to wait for the launch of RXTE
  \citep{Stro97a}.  Surprisingly the frequency of the observed
  oscillations varied during the burst which \citet{Stro97b} argue is a
  hallmark of the conservation of angular momentum during the expansion
  and contraction of the atmosphere.

  The detailed models of \citet{2000ApJ...544..453C} estimate the radius
  expansion expected during the burst.  \citet{Heyl00typei} calculated
  the general relativistic corrections required to translate the
  predictions of \citet{2000ApJ...544..453C} into observable quantities,
  and found that the observed frequency shift would be 30\% - 50\% of
  that predicted by \citet{2000ApJ...544..453C}.  Although
  \citet{2001A&A...374L..16A} \citep[and later][]{2002ApJ...564..343C} found a
  error in the derivation by \citet{Heyl00typei}, Abramowicz and
  colleages also found that the general relativistic decrement is
  large. \citet{2002ApJ...564..343C} also discovered an error in their earlier
  work \citep{2000ApJ...544..453C} which lead them to overestimate the
  frequency shift by a factor of two even in the Newtonian context.

  \citet{2002ApJ...564..343C} pointed out that the frequency shifts
  observed by \citet{2001ApJ...549L..71W} and
  \citet{2001ApJ...549L..85G} were too large to be accounted for by the
  radius expansion models of \citet{Heyl00typei} and
  \citet{2000ApJ...544..453C}.  This letter examines an alternate model
  for both the observed frequencies during Type-I bursts and their
  evolution \citep[first suggested by][]{2000ApJ...544..453C}.  The
  burst may excite modes in the neutron star ocean or photosphere that
  exhibit themselves as dark and light regions on the surface.  Because
  these regions are associated with ocean and atmospheric waves, they
  move relative to the surface of the star.  The properties of the
  atmosphere change as the burst subsides, so the natural frequency of
  the modes shifts accounting for the observed frequency shifts.

  The following section of this letter identifies the modes excited
  by the burst and estimates the observational footprint of these modes.
  The last section examines the implications of these results.

  \section{Waves}

  Neutron stars exhibit several modes of oscillation
  \citep{1987ApJ...318..278M,1996A&A...311..155L}. The modes
  that may be excited by the motion of the deflagation front across the
  surface of neutron star undergoing a Type-I X-ray burst and may explain
  the observations of the burst oscillations share the following 
  features.
  \begin{itemize}
  \item
  {\bf They have a period on the order of one second.}  The observations find that
  the observed frequency shift is a few cycles per second.  
  A mode frequency ($\omega$) of several Hertz is well matched to
  the timescale for the deflagation of the accumulated fuel to envelope
  the stars $\sim 1$~s \citep{1997ApJ...487L..77S}, so it is unlikely
  that much slower modes would be excited. 
  The spin frequencies of the
  stars ($\Omega$) are
  several hundred Hertz so $q=2\Omega/\omega \sim 10^2$.
  \item 
  {\bf The modes should travel westward.}  Here,
  $m>0$ denotes a westbound mode.
  The observed frequency is assumed to be slightly less than the 
  spin frequency of the star, and it increases as the observed photon 
  spectrum cools and dims; consequently, the modes should travel westward, 
  \ie in the opposite sense of the star's rotation.  Some observations have
  found the opposite trend during a portion of the burst, so eastbound 
  modes may also be interesting.
  \item
  {\bf The mode should have no latitudinal nodes,} or  $l_\mu=0$.
  The modes of a rotating star are squeezed near the equator.  Except for 
  special geometries, a band both above and below the equator is visible 
  throughout the star's rotation, so if the mode has latitudinal modes, much of the 
  variability will be averaged out. 
  \item 
  {\bf The azimuthal eigenvalue ($m$) should be 1 or -1.}  Observations have 
  generally found that observed oscillation has a frequency of approximately 
  the spin frequency of the star.  \citet{1999ApJ...515L..77M} 
  found that 4U~1636-536 may be exceptional, so the $|m|=2$ case will also 
  be discussed.
  \item
  {\bf The main observed mode should have no radial nodes in the radial
  displacement,} or $n=1$ ($n$ counts the number of radial nodes in the
  transverse displacement). This ensures that any modes with a similar
  angular dependence will be well separated in frequency, supporting the
  argument that only a particular mode is excited and observed.
  \end{itemize}

  Three prime candidates are the $g-$modes, the buoyant $r-$modes and
  the Kelvin modes, of the neutron star ocean and upper atmosphere.  In
  a rotating neutron star, the modes with various values of numbers of
  azimuthal nodes ($m$) are not degenerate in frequency.  The frequency
  of the modes depends strongly on whether the electrons contribute to
  the entropy of the gas.  If they are degenerate, the
  Brunt-V\"ais\"al\"a frequency ($N$) is reduced by the ratio of the ion
  pressure to the total pressure; however, this reduction is mitigated
  by a corresponding increase in the pressure-scale height ($H$).  
  A first approximation to the frequency of a surface wave is
  \be
  f \sim \frac{N}{2\pi} \frac{H}{R}.
  \ee
  I shall examine modes in a semi-degenerate material with varying
  contributions of ion pressure to the total pressure $\alpha\equiv
  P_i/P_\rmscr{total}$ and non-degenerate material with varying
  contributions of gas and radiation pressure, $\beta\equiv
  P_\rmscr{gas}/P_\rmscr{total}$.  The heat transfer is assumed to 
  be dominated by photons scattered by electrons in the nondegenerate 
  regime and by electrons scattered by ions in the degenerate regime.

  The results of \citet{1996ApJ...460..827B} provide a touchstone for
  the various results.  Extending their results to a semidegenerate
  regime yields the following estimates for the frequency of the modes
  in the rotating frame of the surface of the star,
  \be 
  f_{\lambda,n,\rmscr{D}} \equiv
  \frac{\omega}{2\pi} = 2.37 \rmmat{Hz} \left ( 2 \lambda
  \frac{T}{10^{8}~\rmmat{K}} \frac{56}{A} 
  \frac{8 - 4\alpha - \alpha^2}{16 - \alpha^2}
  \right )^{1/2} \left (
  \frac{10~\rmmat{km}}{R} \right ) \left \{ 1 + n^2 \left [
  \frac{3\pi}{2\ln(\rho_b/\rho_t)} \right ]^2 \right \}^{-1/2}
  \label{eq:freq1D}.
  \ee 
  Here $A$ is the mean atomic weight of non-degenerate species (here,
  the nuclei in the ocean), $\rho_b$ and $\rho_t$ are the densities at
  the top and bottom of the excited layer \citep[][give the dependence
  on ocean depth]{1995ApJ...449..800B}, and $n>0$ is the number of
  radial nodes.  If one takes the limit as $\alpha$ vanishes, the
  \citet{1996ApJ...460..827B} result obtains, and as $\alpha$ approaches
  unity for a fixed value of $A$ the frequency estimate actually
  decreases slightly (by about 40\%).   The frequency of the g-mode 
  is proportional to the Brunt

  The ocean lies below the burning layers from $\rho \sim
  10^7$~g~cm$^{-3}$ to $\rho \sim 10^8$~g~cm$^{-3}$, yielding 4.2 for
  the coefficient of $n^2$ is Eq.~\ref{eq:freq1D}.  The frequency of the
  first harnomic is about half that of the fundamental.  The depth of
  the ocean depends sensitively on the charge of the nuclei produced by
  the burning.  The unstable burning during Type-I bursts produces
  mainly iron group elements \citep{1993SSRv...62..223L}, so the ocean
  is dramatically more shallow than in the higher accretion rate {\it Z}
  sources where the ocean consists mainly of {\it CNO} elements and
  extends to $\rho \sim 10^{11}$~g~cm$^{-3}$
  \citep{1995ApJ...449..800B}.

  In a nondegerate layer, the mode frequencies are several times larger,
  \be
  f_{\lambda,n,\rmscr{ND}} = 14.48 \rmmat{Hz} 
  \left ( 5 \frac{\lambda}{\beta} \frac{T}{10^{8}~\rmmat{K}} \frac{0.6}{\mu} 
  \frac{ 4 - 3\beta }{ 8 - 3\beta } \right )^{1/2}
  \left ( \frac{10~\rmmat{km}}{R} \right ) 
  \left \{ 1 + n^2 \left [ \frac{3\pi}{2\ln(\rho_b/\rho_t)} \right ]^2 \right \}^{-1/2}.
  \label{eq:freq1ND}
  \ee 
  where $\mu$ is the mean molecular weight of the material (this
  includes the electrons because in the non-degenerate case they
  contribute to both the entropy and the pressure) and $\beta$ is the
  ratio of the gas pressure to the total pressure.  To obtain these
  expressions, thermal bouyancy is assumed to dominate
  \citep{2000ApJ...544..453C} and the expressions of
  \citet{1995ApJ...449..800B} and \citet{1996ApJ...460..827B} have been
  scaled by the ratio of the Brunt-V\"ais\"al\"a frequency in the
  nondengerate atmosphere \citep{Bild98} to the degenerate ocean.  If
  one takes $\beta\rightarrow 1$ and $\alpha\rightarrow 1$ in
  expressions Eq.~\ref{eq:freq1D} and~\ref{eq:freq1ND} for the same
  composition, the two expressions agree.  The increase in the frequency
  estimates results entirely from the decrease in the mean molecular
  weight of the nondegenerate species, because the electrons are no longer 
  degenerate. 

  Both equations~\ref{eq:freq1D} and~\ref{eq:freq1ND} include a
  logarithmic factor involving the density of the top and bottom of the
  layers.  \citet{1996ApJ...460..827B} derived this factor when they
  considered a degenerate layer.  It is unclear whether it is indeed
  appropriate in a non-degenerate layer.  This factor changes the
  estimate of the mode frequencies by at most a factor of two for the 
  situations considered here.  Furthermore, in
  the non-degenerate regime the mode period is shorter than the thermal
  time for the layer, so one would expect the modes to be isothermal
  rather than adiabatic.  \citet{1995ApJ...449..800B} examine this point
  in their appendix and find that the lack of adiabaticity changes the
  structure of the mode and most importantly the assumed energy in a
  photospheric mode.  However, they did not determine how it would
  change the frequency of a mode restricted to an isothermal region.

  Unless the temperature is small or $n$ is large, the frequency of the
  oscillation will be much larger than one Hertz.  The latter
  possibility is disfavored according to the criteria listed above.
  Furthermore, to meet the requirements above, $\lambda \sim 1$.

  \citet{Long68} derives the mode structure for the ocean of a rotating
  fluid shell which includes some modes not discussed by 
  \citet{1996ApJ...460..827B}.
  As $q$ increases, for the $g-$modes, $\lambda$ approaches
  \be
  \lambda \sim
  (2 \nu + 1)^2 q^2  \hspace{0.5in} (m>0, \nu = 0, 1,2,\ldots).
  \ee
  where $\nu \geq 1$ is the number of latitudinal nodes in the 
  pertubation to the northward velocity of the fluid. $l_\mu=\nu+1$ for 
  $g$-modes.

  \citet{1996ApJ...460..827B} also discuss a set of eastbound modes which 
  \citet{Long68} identifies as the Kelvin modes with
  \be
  \lambda \sim m^2  \hspace{0.5in} (m<0, \nu = 1)
  \label{eq:lambdakelvin}
  \ee
  which have $l_{\mu}=0$.

  For the buoyant $r-$modes (which all travel westward),
  $\lambda$ approaches,
  \be
  \lambda \sim m^2 (2 \nu + 1)^{-2} \hspace{0.5in} (m>0, \nu = 1,2,\ldots).
  \ee
  and $l_{\mu}=\nu-1$ for the $r$-modes.

  As $\epsilon = q^2 \lambda$ increases the modes become more and more
  localized near the stellar equator, and the eigenfunctions resemble
  parabolic cylinder functions which may be expressed as the product of
  a Hermite polynomial and a Gaussian.  The vertical displacement (or
  equivalently the pressure pertubation) for the mode $\zeta$ is given
  by the following expressions as $\epsilon$ goes to infinity
  \citep{Long68},
  \ba
  \zeta_{g-\rmscr{mode}} &\sim& \frac{1}{(2\nu+1)^{1/2} \epsilon^{1/2}} 
  e^{-\frac{1}{2} \eta^2} \left ( \nu  H_{\nu-1} (\eta) - \frac{1}{2}  H_{\nu+1} (\eta) \right ) e^{m\phi + \omega t}
  \\
  \zeta_\rmscr{Kelvin} &\sim& 
  \frac{2}{m} \epsilon^{1/4}  
  e^{-\frac{1}{2} \eta^2} e^{m\phi + \omega t} \\
  \zeta_{r-\rmscr{mode}} 
  &\sim& \frac{2\nu+1}{2 m} \epsilon^{-1/4} 
  e^{-\frac{1}{2} \eta^2} \left ( H_{\nu-1} (\eta) + \frac{1}{2\nu+2}  
  H_{\nu+1} (\eta) \right ) e^{m\phi + \omega t}
  \ea
  where $\eta=\epsilon^{1/4} \mu$ and $\mu=\cos\theta$.  Most of the
  excitation in a mode lies between $-\sqrt{\nu} < \eta < \sqrt{\nu}$.
  For the different modes, this leads to the following range in $\mu$,
  \be
  |\mu| < \left \{ 
  \begin{array}{ll} 
  q^{-1} \left ( 2 + \nu^{-1} \right )^{-1/2} & \rmmat{$g$-modes} \\
  q^{-1/2} |m|^{-1/2}  & \rmmat{Kelvin modes.} \\
  q^{-1/2} |m|^{-1/2} 
  \left ( 2 \nu^2 + \nu \right )^{1/2} & \rmmat{buoyant $r$-modes} 
  \end{array} 
  \right .
  \ee
  The buoyant $r$-modes occupy the widest band near the equator; the
  variability is approximately proportional to the square root of the
  width of the band, so the $r-$modes should exhibit the largest variability.

  The next subsection focuses on the modes which are likely to be
  excited in the ocean and what their observational footprint would be.
  The modes tend to become more localized in the equatorial regions as
  the spin of the star increases \citep{1996ApJ...460..827B}.  The
  observations indicate that the excited modes typically have $|m|=1$
  \citep{Stro97b} or possibly $|m|=2$ \citep{1999ApJ...515L..77M}.

  The frequency of the modes in the observer's frame is given by
  $\omega_I = \omega - m \Omega$ (neglecting the gravitational redshift
  for now), so for $m>0$ one would see the frequency of the variation
  increase as the temperature of the ocean decreases.  This is what is
  usually observed \citep{2000ApJ...544..453C}; therefore, the detailed
  calculations will focus on the $m=1$ and $m=2$ modes with frequencies
  of about one cycle per second.

  \subsection{Visibility}

  Calculating the bolometric pulsed fraction for the mode provides an
  estimate the visibility of the various modes.  Because the frequency of
  the mode is about one Hertz and the rotational frequency of the star
  is about 300~Hz, a value of $q=300$ is appropriate to calculate the
  modes.  The variability of these modes decreases as $q^{-1/4}$.  The
  observer lies at a latitude of $30^\circ$ and compare the total flux
  observed when the observer lies above a bright patch (intensity
  maximum) patch to when the observer lies above a faint patch (where
  the intensity vanishes).  The pulsed fraction so defined is
  \be 
  \rmmat{PF} = \frac{f(\rmmat{bright}) - f(\rmmat{faint})} 
		    {f(\rmmat{bright}) + f(\rmmat{faint})} 
  \ee 
  and can range from $-1$ to $1$.  The pulsed fraction is negative if
  the total flux observed from the star is larger when the observer lies
  above a faint region.

  Lacking a specific model for how the presense of a mode induces
  variations in the flux, the effective temperature of each surface
  element is assumed to vary as the linear combination of a constant and
  the vertical displacement induced by the mode of interest.  The
  normalization of the mode is selected so that the effective
  temperature varies over the surface from zero to twice the underlying
  value.  Because the observed variability is small during the tail 
  of the burst, the particular choice of coupling is unimportant; one
  can always choose a different type of connection between the mode and
  the observed flux and rescale the size of the underlying wave. 

\begin{figure}
  {\plottwo{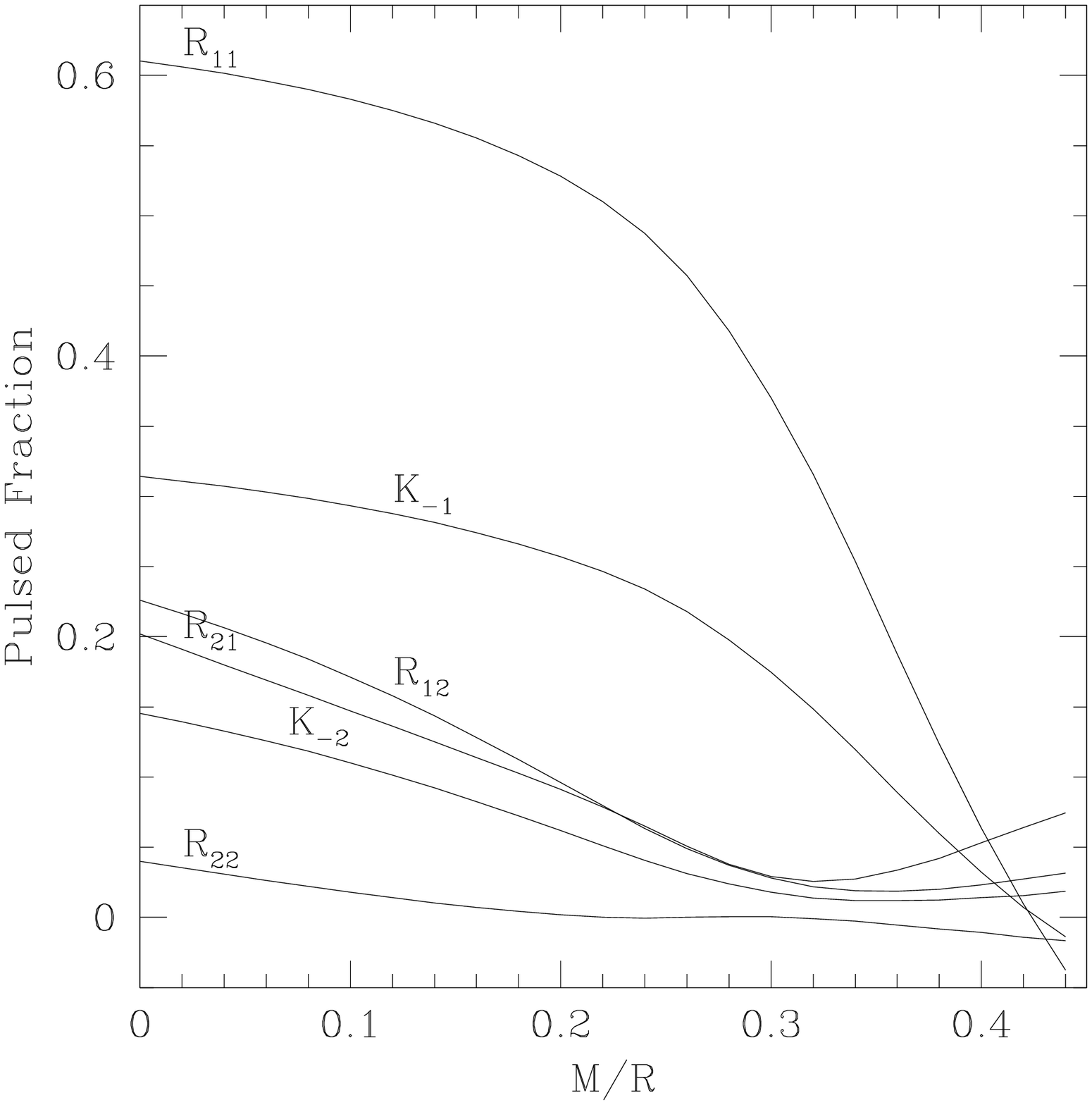}{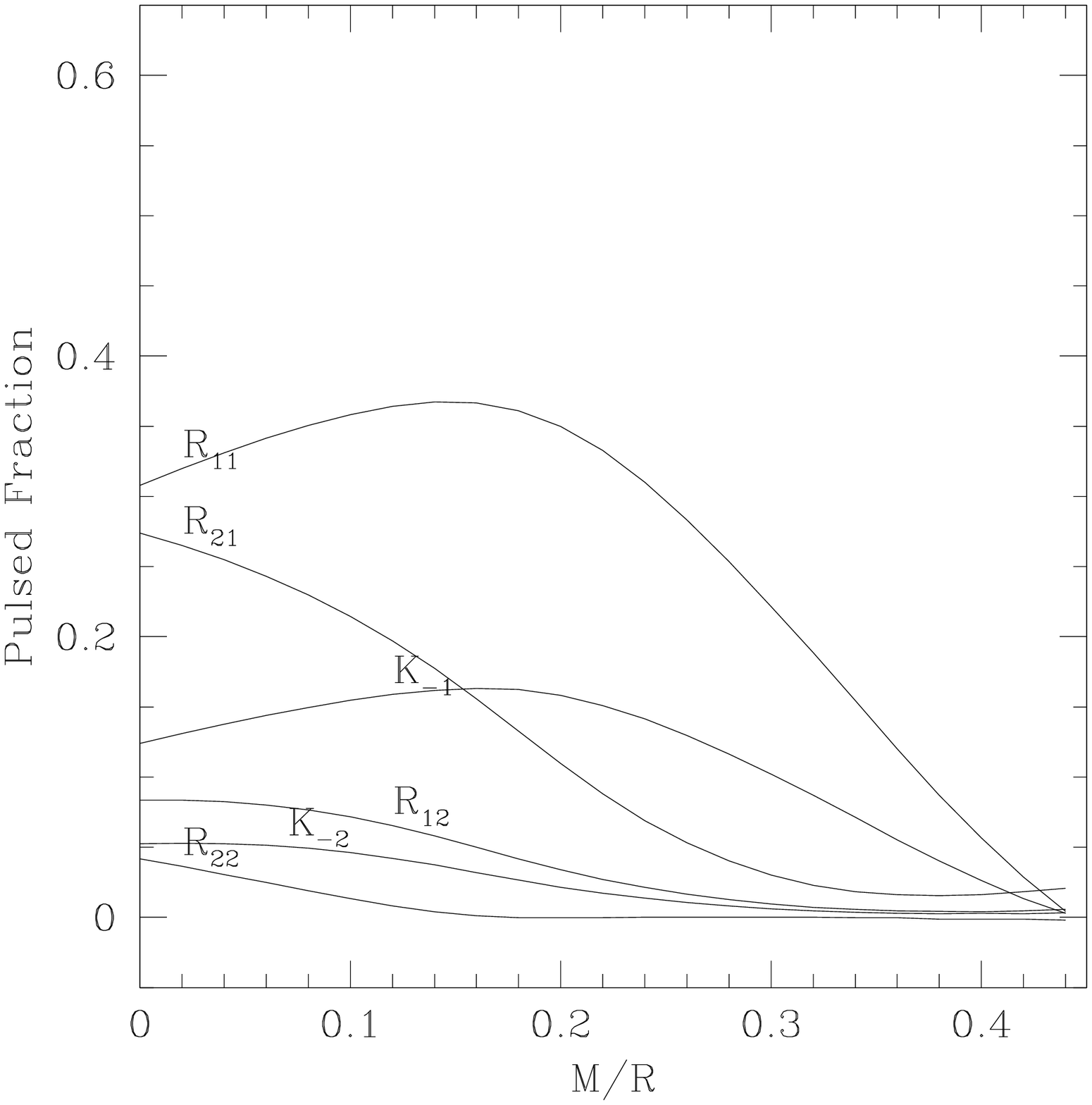}}

  {\plottwo{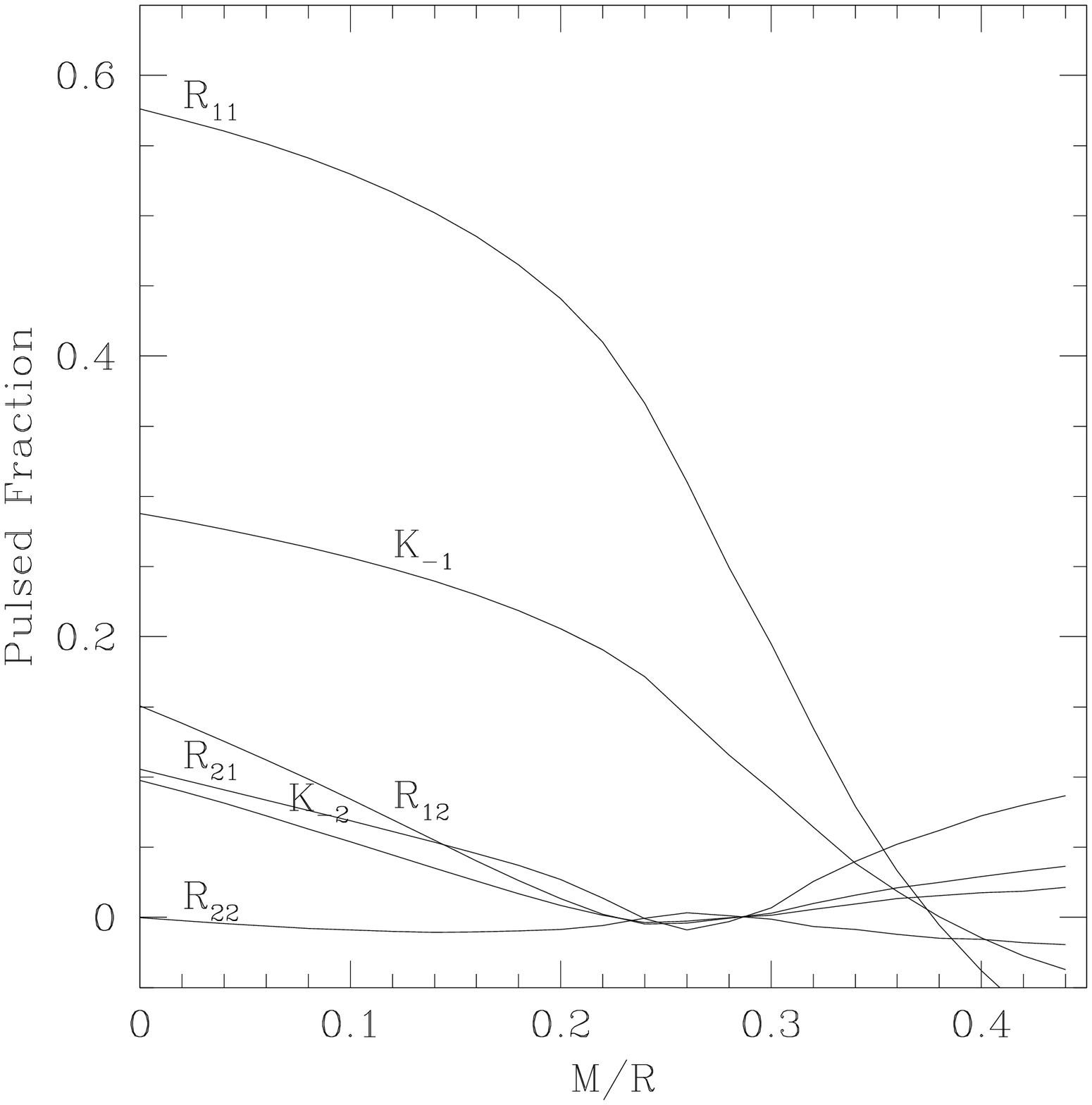}{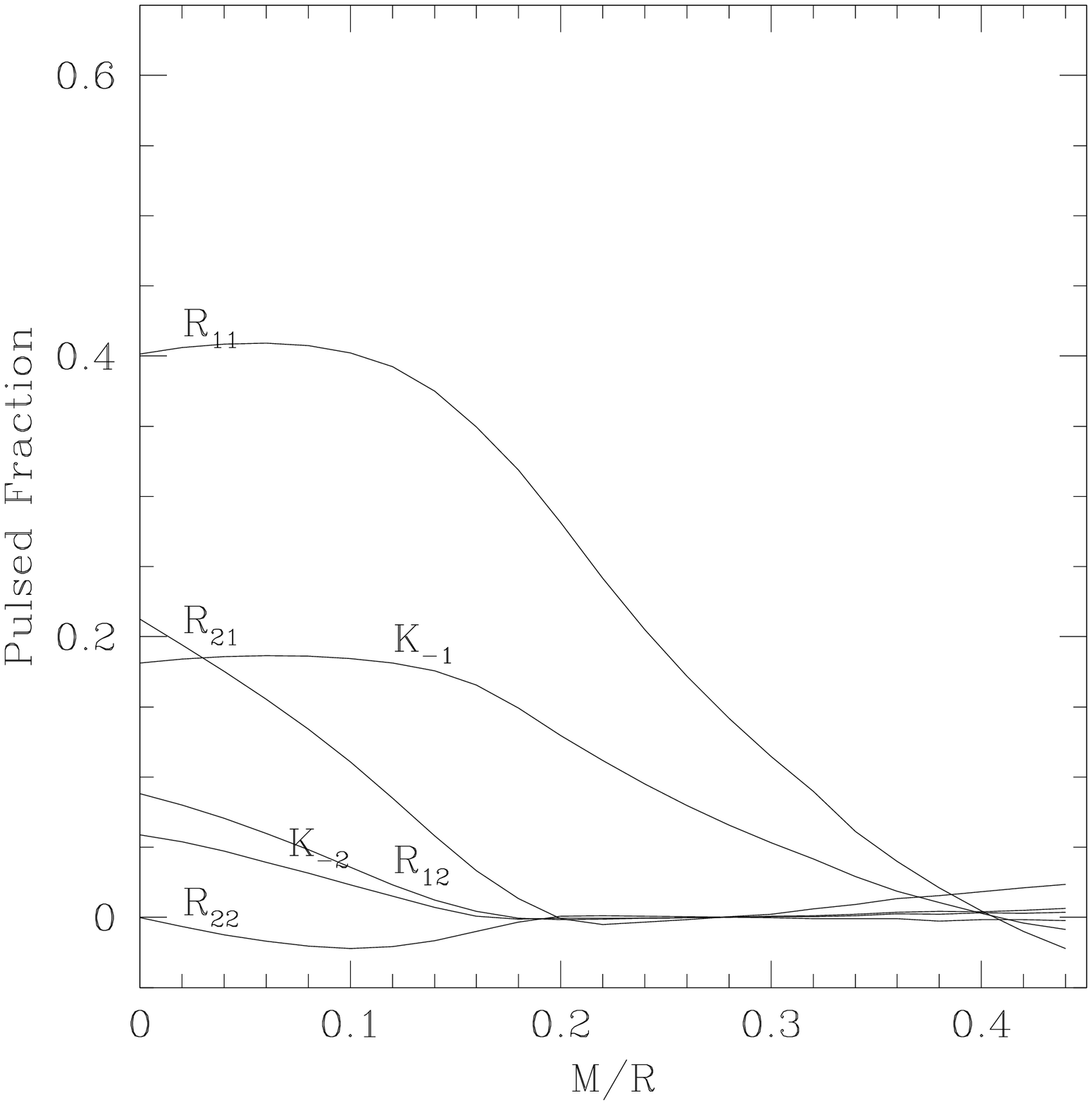}}
  \caption{Pulsed Fraction for various modes as a function of $M/R$ for 
  an observer latitude of $30^\circ$ (left) and $60^{\circ}$ (right).  
  The upper curves include a rudmimentary treatment of limb darkening. A 
  negative pulsed fraction indicates that the star is brightest when the 
  observer is directly above a dark region. $R_{\nu m}$ denotes the
  Rossby mode with the particular value of $\nu$ and $m$.  $K_{m}$
  denotes a Kelvin mode.}
  \label{fig:pf}
  \end{figure}

  The gravitiational defocussing of the neutron star surface in
  the Schwarschild geometry \citep[\eg][]{Heyl97analns,Page95} and limb
  darkening also affect the variability.  The limb darkening is modeled
  by allowing the intensity from a given surface element to vary as the
  cosine of the zenith angle \citep[\eg][]{Pern00pulsar}.  \figref{pf}
  presents the results for the first few modes with $m=1$ and $m=2$.
  The inclusion of limb darkening increases the expected variability
  significantly and also alleviates the effects of gravitational
  defocussing.

\clearpage

\clearpage

  As expected from the earlier discussion, the buoyant $r-$modes exhibit
  the largest variability.  Modes with a larger number of nodes $\nu$
  show less variability as do modes with $|m|>1$.  The Kelvin modes are
  more concentrated near the equator, and therefore exhibit less
  variability.  The $g-$modes which are extremely concentrated near the
  equator show negligible variability when averaged over the visible
  portion of the star with a static background.  Of all the modes in the
  ocean for a given excitation, the lowest-order $r-$mode exhibits the
  largest variability and also satisfies the observational constraints.
  The absolute variability of a particular mode depends on the {\it
  local} coupling with the mode and the emergent flux; the details of
  this coupling are beyond the scope of this paper, but the relative
  variability between the different modes will not depend on these
  details as long as the coupling is local.

  \subsection{Variability Frequency}

  With a mode excited, a distant observer would find that the flux from
  the star would vary with a frequency of $m\Omega-\omega$, slightly
  less than the spin frequency of the star for $m=1$.  

  \subsubsection{A Degnerate Layer}

  If the mode resides in the degenerate ocean lying beneath the burning
  layers, the frequency of
  the mode is given by Eq. 15 of \citet{1996ApJ...460..827B}
  or Eq.~\ref{eq:freq1D} as $\alpha\rightarrow 0$,
  \be
  f_{\lambda,n,\rmscr{D}} \equiv \frac{\omega}{2\pi} = 2.37 \rmmat{Hz} 
  \left ( m^2 \frac{T}{10^{8}~\rmmat{K}} \frac{56}{A} \right )^{1/2}
  \left ( \frac{10~\rmmat{km}}{R} \right ) 
  \left ( 1 + 4.2 n^2 \right )^{-1/2} 
  \times \left \{
  \begin{array}{ll}
  \left ( 2 \nu + 1 \right )^{-1} & \rmmat{$r$-modes} \\
  1                        & \rmmat{Kelvin modes}
  \end{array}
  \right .
  \label{eq:freq2D}
  \ee 
  in the limit of rapid rotation, \ie large $q=2\Omega/\omega$.  The
  ocean is left after a Type-I burst which burns the accreted hydrogen
  and helium directly to the iron group elements
  \citep{1993SSRv...62..223L} and possibly beyond
  \citep{2001PhRvL..86.3471S}; for simplicity $A$ is set at 56 ;
  \citet{1996ApJ...460..827B} were concerned with the carbon ocean
  remaining after the steady nuclear burning associated with the more
  rapidly accreting ``Z'' sources.  This frequency estimate assumes that
  the material in the excited layer is degenerate while the mode is
  excited.

  The temperature at the bottom of the burning layers at the peak of the
  burst is typically around $10^{9}$~K and drops to $10^{8}$~K as the
  burst subsides \citep[e.g.][]{Joss78}.  For the $\nu=1$ and $n=1$
  westbound $r-$mode in the burning layer itself, the
  degenerate formula (Eq.\ref{eq:freq2D} yields a frequency shift from
  $1.62$~Hz at the peak temperature of $2.2\times 10^9$~K 
  \citep{2000ApJ...544..453C} to $0.35$~Hz at $10^8$~K.  Even
  this modest frequency shift would require that the massive
  ocean be well coupled to the burning layers above.
  Because the mass of the ocean is much greater than that of the
  burning layers, it is unlikely that the ingoing flux during the
  burst is sufficient to heat up the ocean to $10^9$~K.

  Modes with higher values of $n$ will exhibit smaller frequency shifts
  --- for $n=2$ the shift is 45\% smaller.  While modes with $m<-1$ will
  have larger shifts (but the variability will be found near $|m|$ times
  the rotational frequency), the observed variation in the total flux is
  less.  The Kelvin mode exhibits a frequency shift that is three times
  larger because $\lambda$ is nine times larger (see
  Eq.~\ref{eq:lambdakelvin}).  The observed frequency of the mode
  decreases as the ocean cools.

  \subsubsection{A Non-degnerate Layer}

  The burning layers become sufficiently hot to lift the electron
  degeneracy \citep{2000ApJ...544..453C}, the frequency shift would be
  an order of magnitude larger.  Additionally, because I have assumed
  that electron scattering dominates the opacity, the ratio of the gas
  pressure to the total pressure ($\beta$) is constant through the
  non-degenerate layer where neither convection nor nuclear burning
  occur.  If I take $\beta$ equal to unity, and the temperature of the
  burning layer going from $10^9$~K to $10^8$~K while the mode is
  excited, the mode frequency will change by about eight Hertz much
  larger than is observed.

  However, a mode could naturally become trapped in the photosphere
  itself where the gas is typically much cooler.
  \citet{1994ApJ...431L.103L} argued that sound waves can be trapped
  above the photosphere of the neutron star.  Mdoes are trapped above
  the photosphere of the sun in the chromosphere between the temperature
  minimum and the steep temperature gradient at higher altitudes
  \cite[e.g.][]{1982ApJ...258..393L} It is natural to speculate that
  something similar could occur in the atmosphere of a Type-I burst.
  \citet{1994ApJ...431L.103L} considered situations where $\beta\ll\ 1$
  and the photosphere expanded to several stellar radii (\ie radius
  expansion bursts).  Here I will focus on gravity waves in the
  relatively thin photospheres of burst that do not experience radius
  expansion.

  Because $\beta$ is constant through the photosphere and it
  is potentially observable, it is natural to use the condition of
  hydrostatic equilibrium to eliminate the temperature from the
  frequency estimate.
  \begin{equation}
  T_\rmscr{eff}^4 = (1-\beta) \frac{c g_s}{\sigma \kappa} =
  (1-\beta)
  \frac{g_s}{2.4\times 10^{14} \rmmat{cm s}^{-2}}
  \left ( 2.4 \times 10^7 \rmmat{K} \right )^4
  \rmmat{~and~} 1-\beta = \frac{F}{F_\rmscr{Edd}}
  \label{eq:photodef}
  \end{equation}
  This substitution in Eq.~\ref{eq:freq1ND} yields
  \begin{eqnarray}
  f_{\lambda,n,\rmscr{ND}} &=& 7.09 \rmmat{Hz} 
  \left ( 5 m^2 \frac{(1-\beta)^{1/4}}{\beta} \frac{ 4 - 3\beta }{ 8 - 3\beta }
  \frac{0.6}{\mu} \right )^{1/2}
  \left ( 1 + 4.2 n^2 \right )^{-1/2} \times
  \nonumber \\
  & & ~~~~
  \left ( \frac{g_s}{2.4\times 10^{14} \rmmat{cm s}^{-2}} \right )^{1/8}
  \left ( \frac{10~\rmmat{km}}{R} \right ) 
  \times \left \{
  \begin{array}{ll}
  \left ( 2 \nu + 1 \right )^{-1} & \rmmat{$r$-modes} \\
  1                        & \rmmat{Kelvin modes}
  \end{array}
  \right .
  \label{eq:freq2ND}
  \end{eqnarray}
  where I have assumed that the photosphere spans an order of magnitude 
  in density. 
  Two important features of Eq.~\ref{eq:freq2ND} is that it depends on 
  observable quantities and that the frequency drift diverges as the flux 
  approaches the Eddington limit.

  If the burst oscillation depicted in Fig. 2 of
  \citet{1999ApJ...516L..81S} is taken as an example with the assumption
  that the burst peaked very close to the Eddington flux, the
  oscillation starts with $F/F_\rmscr{Edd} \approx 0.9$ and continues
  until $F/F_\rmscr{Edd} \approx 0.1$, which yields a frequency shift of
  4.1~Hz (or 3.07~Hz including a gravitational redshift of 0.35) about
  twice the observed value.  \citet{1999ApJ...523L..51S} found
  oscillations whose frequency increased as the burst subsides.  The
  excitation of a Kelvin mode could naturally explain this behavior.
  The spin-down portion of the oscillation begins when the flux is 0.6
  of its peak value and continues until the fraction is 0.3 and its
  frequency increases by 1.4~Hz.  The model predicts a slightly larger
  shift of 1.9~Hz (again with the gravitational redshift).

  It is crucial to emphasize that the estimates of the frequency shifts
  given in the previous paragraph are upper bounds for a layer of a
  given thickness in density.  If the composition of the photosphere is
  not solar but helium, the frequency shifts would decrease by
  one-third.  Less subtle is the assumption that the bursts reach the
  Eddington rate at their peak.  For example if the two example bursts
  discussed earlier peaked at 80\% of the Eddington rate the frequency
  shifts would be 1.43~Hz for the spin-up example -- almost exactly the
  value found by \citet{1999ApJ...516L..81S} -- and 1.30~Hz for the
  spin-down case.  In sources with radius expansion bursts, the peak
  burst flux can be calibrated and these predicted trends tested in
  detail to possibly give hints of the radii and gravitational redshifts
  of the underlying neutron stars.

  The observed burst oscillations have been found to have $Q-$values of
  several thousand.  Since the frequency of the mode is boosted by a
  factor of several hundred, this is consistent with the mode having a
  $Q-$value on the order of ten.  The observed value of $Q$ is given by
  the product of the ratio of the observed frequency to the inherent
  frequency of the mode and the $Q-$value of the mode.  Such moderate
  values of $Q$ are not difficult to achieve in the neutron star ocean
  \citep{1995ApJ...449..800B}.  The oscillations of interest suffer
  little dissipation on the timescale of the bursts.  These details of
  the photospheric modes have not yet been determined.

  \section{Discussion}

  The preceding sections have presented a model for the observed
  varability in Type-I bursts and its shift in frequency.  Unlike the
  simple rotational modulation model, the excitation and evolution of
  $r$-modes in the neutron-star ocean can naturally account for the size
  and sign of the shift and the presence of variability even after the
  burst has enveloped the entire star.

  During the onset of the burst when only a portion of the star is hot,
  it is quite natural to account for the variability as a growing
  hotspot \citep{1997ApJ...487L..77S}.  \citet{1998ApJ...498L.135S}
  found that the pulsed fraction of 75$\pm$15\% during the onset of the
  burst.  The presence and growth of this hotspot sets up travelling
  modes in the ocean.  Due to a match in timescales between the growth
  of the spot and the oscillation of the modes, modes with frequencies
  near a Hertz are preferentially excited.  Modes with $m=1$ result in
  the largest observed variability.  From the observer's point of view,
  the flux varies at slightly less than the rotational frequency of the
  star.  During the decay of the burst, the observed modulation is
  significantly lower $\sim 15\%$ \citep{Stro98}.  The $r-$modes with
  the simple coupling considered in the letter can easily generate this
  amplitude of modulation.

  Although modes with higher (more negative) values of $m$ may be
  excited, the resulting variability is much smaller.  Since the burst
  begins in a particular spot and grows, one would expect modes with odd
  values of $m$ to be excited preferentially.  The resulting variability
  for the $m=3$ mode is a factor of 30--100 smaller than the $m=1$ mode;
  this is well below the upper limits quoted by \citet{2002ApJ...581..550M}
  for several sources.

  \citet{1999ApJ...515L..77M} discovered that 4U~1636-536 exhibits
  oscillations at 580~Hz as well as much weaker oscillations 290~Hz; on
  the basis of other observations (QPOs) and a model of the accretion
  flow \citep{1998ApJ...508..791M}, Miller argues that the latter is the
  spin frequency of the star and the former is its first overtone.  He
  associates the 580~Hz signal with the presence of two nearly antipodal
  hotspots burning on the surface \citep{2001ApJ...546.1098W}.  Recent
  observations of burst oscillations from SAX~J1808.4-3658 at the known
  spin frequency of about 400~Hz clarifies this issue
  \citep{2003HEAD...35.4501C}.  The difference in the frequencies of the
  QPOs is only about 200~Hz, so in other sources where the burst
  oscillation frequency is twice the QPO difference, the burst frequency
  may also be the spin frequency of the star, so the $|m|=1$ mode is
  indeed the correct one to consider.

  On the other hand, if 290~Hz is indeed the spin frequency of
  4U~1636-536, in these particular bursts, anomalously, the $m=2$ mode
  is excited more strongly that the $m=1$ mode.  If both modes have the
  same value of $n$ (the number of radial nodes), a prediction of the
  model would be that the frequency shift would be twice as large for
  the 580~Hz variation as for the 290~Hz variation.  The simple
  rotational modulation model predicts a similar trend.  However, the
  value of $\omega$ could be chosen by the evolution of the burning
  front, so the two modes could correspond to different values of $n$
  with the nearly same value of $\omega$.  In this case, both
  oscillations would vary in frequency by the same amount.

  This brings up the important question of why the burst only excites a
  mode with a particular number of radial nodes.  If this were not the
  case, power would appear at several frequencies near the spin
  frequency of the star and these frequencies would all evolve according
  to \eqref{freq1ND}.  Perhaps it is not surprising that only a mode
  with a $n=1$ is excited since the excitation of higher modes requires
  additional shearing in the middle of the excited layer.  The burst is
  associated with activity only in a thin layer at the top of ocean or at
  the bottom of the photosphere.

  This model makes several predictions.  In principle eastbound modes
  may be excited.  This would result in an observed frequency decrease
  as the atmosphere cools after the burst about three times larger than
  that caused by the westbound modes
  \citep[e.g.][]{1999ApJ...523L..51S}.  More generally, several modes
  could be excited.  If several modes were excited, one could in
  principle derive properties of the neutron star since the spacing of
  the modes is well understood.  These higher order modes would
  typically exhibit less variability since the variability is
  proportional to $q^{-1/4}$ or $\omega^{1/4}$.

  Furthermore, since the frequency of the mode changes by a factor of
  several as the surface layers cool, one would expect the variability
  to decrease slightly during this epoch as well.  Additionally, more
  quickly rotating stars should exhibit slightly less variability and
  more importantly similar frequency shifts, \ie the frequency shift is
  not proportional to $\Omega$ in contrast with standard,
  angular-momentum conservation model.  \citet{2000ApJ...544..453C}
  noted that the frequency shift rather than the fractional frequency
  shift is similar from source to source.  Also, because the mode may
  still be excited after the ocean cools to its equilibrium temperature,
  one need not expect the observed frequency of the oscillation to
  asymptotically approach the precisely same value from one burst to the
  next.

  \citet{2002ApJ...580.1048M} noted that the asymptotic frequency is
  stable to a few parts per thousand from burst to burst.  For the
  photospheric model, the limit of the observed frequency as the burst
  flux goes to zero is indeed the spin rate of the underlying neutron
  star and should be constant from burst to burst.
  \citet{2002ApJ...580.1048M} also found that a small fraction of bursts
  exhibited multiple oscillation frequencies or spin-down episodes which
  this model exhibits.  Most interestingly they found that the burst
  oscillations cease during radius expansion episodes.  During a radius
  expansion episode the frequency of the photospheric modes
  diverges so the photospheric model argues that no oscillations should
  be found during radius expansion.

  \citet{2000ApJ...544..453C} argued that even a small magnetic field
  could play a dynamic role in the frequency shift of burst
  oscillations.  In their layer model, the shearing fluid could
  dramatically stretch the magnetic-field lines which could provide a
  significant resistance to the shearing motion and limit the expected
  frequency shifts.  In this model, the fluid itself does not move a
  significant distance relative to the stellar surface so the magnetic
  field does not get stretched significantly.  In principle a
  sufficiently strong magnetic field could change the mode frequencies
  \citep[e.g.][see Eq.~1]{2002ApJ...574..908M} but a detailed discussion
  of the effects of weak magnetic fields on these models is beyond the
  scope of these article.

  This letter proposes a model for Type-I burst oscillations in which
  the burst excites waves in either the degenerate ocean, the burning
  layer itself or the photosphere.  The frequency in the rotating frame
  of the waves in the ocean and the photosphere are typically several
  Hertz which provides a good match with the deflagation timescale.
  The frequency is the burning layers is a factor of ten larger.

  Even though the ocean model does yield mode frequencies near the
  observed values, its thermal inertia is large so it is difficult to
  understand how its properties could change so dramatically as the
  burst subsides.  The photospheric model best accounts for the observed
  frequency shifts.  The predicted frequency shift depends on the
  potentially observable value of the ratio of the outgoing flux to the
  Eddington value, and it is close to the observed values.

  However, further work is necessary.  The frequency estimates presented
  here were derived using several simplify assumptions.  First, it was
  assumed that the mode frequencies scale simply with the
  Brunt-V\"ais\"al\"a frequency as one goes from a degenerate to a
  non-degenerate to a radiation-pressure-dominated regime.  Second, the
  adiabatic estimates of \citet{1995ApJ...449..800B} and
  \citet{1996ApJ...460..827B} were used in the photosphere where a more
  realistic isothermal approximation would be appropriate.  Third, the
  magnetic field of the neutron star could also affect the frequency of
  modes in the tenuous photosphere \citep[see][for a
  discussion]{1995ApJ...449..800B}.  A more detailed treatment of the
  physics of the photosphere of a neutron star during a Type-I burst
  could address all of these concerns as well as determine the
  photospheric structure of a neutron star during a Type-I burst and
  would be very worthwhile given the approximate agreement between these
  gravity-wave models and the observations.

  In general these gravity-wave models present the possibility of a
  family of modes being excited during the burst whose observed
  frequencies may increase or decrease as the ocean cools.  Modes with
  more radial, azimuthal and latitudinal nodes tend to exhibit less
  variability when averaged over the observed portion of the stellar
  surface -- the mode which lacks latitudinal and radial nodes and has
  only one azimuthal mode exhibits a significantly larger variability
  than the other modes, independent of viewing angle.  Finally, the
  magnitude of the frequency shift should be independent of the spin
  rate of the star, in contrast to the standard picture of Type-I burst
  oscillations.

  \acknowledgments 
  Support for this work was provided by the National Aeronautics and
  Space Administration through Chandra Postdoctoral Fellowship Award
  Number PF0-10015 issued by the Chandra X-ray Observatory Center, which
  is operated by the Smithsonian Astrophysical Observatory for and on
  behalf of NASA under contract NAS8-39073.  I would like to acknowledge
  useful discussions with Greg Ushomirsky, Avi Loeb and Dimitrios
  Psaltis and helpful comments from the referee.

  \def\mn{MNRAS}


\clearpage


\begin{thebibliography}{42}
\expandafter\ifx\csname natexlab\endcsname\relax\def\natexlab#1{#1}\fi

\bibitem[{{Abramowicz} {et~al.}(2001){Abramowicz}, {Klu{\' z}niak}, \&
  {Lasota}}]{2001A&A...374L..16A}
{Abramowicz}, M.~A., {Klu{\' z}niak}, W., \& {Lasota}, J.~P. 2001, \aap, 374,
  L16

\bibitem[{Bildsten(1998)}]{Bild98}
Bildsten, L. 1998, in The Many Faces of Neutron Stars, ed. A.~Alpar,
  L.~Buccheri, \& J.~van Paradij (Dordrecht: Kluwer), 419, astro-ph/9709094

\bibitem[{{Bildsten} \& {Cutler}(1995)}]{1995ApJ...449..800B}
{Bildsten}, L. \& {Cutler}, C. 1995, \apj, 449, 800+, (BC95)

\bibitem[{{Bildsten} {et~al.}(1996){Bildsten}, {Ushomirsky}, \&
  {Cutler}}]{1996ApJ...460..827B}
{Bildsten}, L., {Ushomirsky}, G., \& {Cutler}, C. 1996, \apj, 460, 827+

\bibitem[{{Chakrabarty} {et~al.}(2003){Chakrabarty}, {Morgan}, {Wijnands}, {van
  der Klis}, {Galloway}, {Muno}, \& {Markwardt}}]{2003HEAD...35.4501C}
{Chakrabarty}, D., {Morgan}, E.~H., {Wijnands}, R., {van der Klis}, M.,
  {Galloway}, D.~K., {Muno}, M.~P., \& {Markwardt}, C.~B. 2003, AAS/High Energy
  Astrophysics Division, 35, 0

\bibitem[{{Cumming} \& {Bildsten}(2000)}]{2000ApJ...544..453C}
{Cumming}, A. \& {Bildsten}, L. 2000, \apj, 544, 453

\bibitem[{{Cumming} {et~al.}(2002){Cumming}, {Morsink}, {Bildsten}, {Friedman},
  \& {Holz}}]{2002ApJ...564..343C}
{Cumming}, A., {Morsink}, S.~M., {Bildsten}, L., {Friedman}, J.~L., \& {Holz},
  D.~E. 2002, \apj, 564, 343

\bibitem[{{Galloway} {et~al.}(2001){Galloway}, {Chakrabarty}, {Muno}, \&
  {Savov}}]{2001ApJ...549L..85G}
{Galloway}, D.~K., {Chakrabarty}, D., {Muno}, M.~P., \& {Savov}, P. 2001,
  \apjl, 549, L85

\bibitem[{{Grindlay} {et~al.}(1976){Grindlay}, {Gursky}, {Schnopper},
  {Parsignault}, {Heise}, {Brinkman}, \& {Schrijver}}]{1976ApJ...205L.127G}
{Grindlay}, J., {Gursky}, H., {Schnopper}, H., {Parsignault}, D.~R., {Heise},
  J., {Brinkman}, A.~C., \& {Schrijver}, J. 1976, \apjl, 205, L127

\bibitem[{{Hansen} \& {van Horn}(1975)}]{1975ApJ...195..735H}
{Hansen}, C.~J. \& {van Horn}, H.~M. 1975, \apj, 195, 735

\bibitem[{Heyl(2000)}]{Heyl00typei}
Heyl, J.~S. 2000, \apjl, 542, 45

\bibitem[{Heyl \& Hernquist(1998)}]{Heyl97analns}
Heyl, J.~S. \& Hernquist, L. 1998, \mn, 300, 599

\bibitem[{{Joss}(1977)}]{1977Natur.270..310J}
{Joss}, P.~C. 1977, \nat, 270, 310

\bibitem[{{Joss}(1978)}]{Joss78}
---. 1978, \apjl, 225, L123

\bibitem[{{Lamb} \& {Lamb}(1978)}]{1978ApJ...220..291L}
{Lamb}, D.~Q. \& {Lamb}, F.~K. 1978, \apj, 220, 291

\bibitem[{{Lapidus} {et~al.}(1994){Lapidus}, {Nobili}, \&
  {Turolla}}]{1994ApJ...431L.103L}
{Lapidus}, I., {Nobili}, L., \& {Turolla}, R. 1994, \apjl, 431, L103

\bibitem[{{Lee} \& {Strohmayer}(1996)}]{1996A&A...311..155L}
{Lee}, U. \& {Strohmayer}, T.~E. 1996, \aap, 311, 155

\bibitem[{{Leibacher} {et~al.}(1982){Leibacher}, {Gouttebroze}, \&
  {Stein}}]{1982ApJ...258..393L}
{Leibacher}, J., {Gouttebroze}, P., \& {Stein}, R.~F. 1982, \apj, 258, 393

\bibitem[{{Lewin} {et~al.}(1993){Lewin}, {van Paradijs}, \&
  {Taam}}]{1993SSRv...62..223L}
{Lewin}, W.~H.~G., {van Paradijs}, J., \& {Taam}, R.~E. 1993, Space Science
  Reviews, 62, 223+

\bibitem[{Longuet-Higgins(1968)}]{Long68}
Longuet-Higgins, M.~S. 1968, Phil. Trans. R. Soc., 262, 511, (LH68)

\bibitem[{{McDermott} \& {Taam}(1987)}]{1987ApJ...318..278M}
{McDermott}, P.~N. \& {Taam}, R.~E. 1987, \apj, 318, 278

\bibitem[{{Miller}(1999)}]{1999ApJ...515L..77M}
{Miller}, M.~C. 1999, \apjl, 515, L77

\bibitem[{{Miller} {et~al.}(1998){Miller}, {Lamb}, \&
  {Psaltis}}]{1998ApJ...508..791M}
{Miller}, M.~C., {Lamb}, F.~K., \& {Psaltis}, D. 1998, \apj, 508, 791

\bibitem[{{Morsink} \& {Rezania}(2002)}]{2002ApJ...574..908M}
{Morsink}, S.~M. \& {Rezania}, V. 2002, \apj, 574, 908

\bibitem[{{Muno} {et~al.}(2002{\natexlab{a}}){Muno}, {{\" O}zel}, \&
  {Chakrabarty}}]{2002ApJ...581..550M}
{Muno}, M.~P., {{\" O}zel}, F., \& {Chakrabarty}, D. 2002{\natexlab{a}}, \apj,
  581, 550

\bibitem[{{Muno} {et~al.}(2002{\natexlab{b}}){Muno}, {Chakrabarty}, {Galloway},
  \& {Psaltis}}]{2002ApJ...580.1048M}
{Muno}, M.~P., {Chakrabarty}, D., {Galloway}, D.~K., \& {Psaltis}, D.
  2002{\natexlab{b}}, \apj, 580, 1048

\bibitem[{Page(1995)}]{Page95}
Page, D. 1995, ApJ, 442, 273

\bibitem[{Perna {et~al.}(2001)Perna, Heyl, \& Hernquist}]{Pern00pulsar}
Perna, R., Heyl, J., \& Hernquist, L. 2001, \apj, 553, 809

\bibitem[{{Schatz} {et~al.}(2001){Schatz}, {Aprahamian}, {Barnard}, {Bildsten},
  {Cumming}, {Ouellette}, {Rauscher}, {Thielemann}, \&
  {Wiescher}}]{2001PhRvL..86.3471S}
{Schatz}, H., {Aprahamian}, A., {Barnard}, V., {Bildsten}, L., {Cumming}, A.,
  {Ouellette}, M., {Rauscher}, T., {Thielemann}, F.-K., \& {Wiescher}, M. 2001,
  Physical Review Letters, 86, 3471

\bibitem[{{Spitkovsky} {et~al.}(2002){Spitkovsky}, {Levin}, \&
  {Ushomirsky}}]{2002ApJ...566.1018S}
{Spitkovsky}, A., {Levin}, Y., \& {Ushomirsky}, G. 2002, \apj, 566, 1018

\bibitem[{Strohmayer \& Bildsten(2003)}]{Stro03}
Strohmayer, T. \& Bildsten, L. 2003, in Compact Stellar X-Ray Sources, ed.
  W.~Lewin \& M.~van~der Klis (Cambridge: Cambridge University Press),
  astro-ph/0301544

\bibitem[{Strohmayer {et~al.}(1997)Strohmayer, Zhang, \& Swank}]{Stro97a}
Strohmayer, T., Zhang, W., \& Swank, J. 1997, \apjl, 487, 77

\bibitem[{{Strohmayer}(1999)}]{1999ApJ...523L..51S}
{Strohmayer}, T.~E. 1999, \apjl, 523, L51

\bibitem[{{Strohmayer} {et~al.}(1997{\natexlab{a}}){Strohmayer}, {Jahoda},
  {Giles}, \& {Lee}}]{Stro97b}
{Strohmayer}, T.~E., {Jahoda}, K., {Giles}, A.~B., \& {Lee}, U.
  1997{\natexlab{a}}, \apj, 486, 355+

\bibitem[{{Strohmayer} \& {Markwardt}(1999)}]{1999ApJ...516L..81S}
{Strohmayer}, T.~E. \& {Markwardt}, C.~B. 1999, \apjl, 516, L81

\bibitem[{Strohmayer {et~al.}(1998)Strohmayer, Swank, \& Zhang}]{Stro98}
Strohmayer, T.~E., Swank, J.~H., \& Zhang, W. 1998, Nucl. Phys. B (Proc.
  Suppl.), 69, 129, astro-ph/9801219

\bibitem[{{Strohmayer} {et~al.}(1997{\natexlab{b}}){Strohmayer}, {Zhang}, \&
  {Swank}}]{1997ApJ...487L..77S}
{Strohmayer}, T.~E., {Zhang}, W., \& {Swank}, J.~H. 1997{\natexlab{b}}, \apjl,
  487, L77

\bibitem[{{Strohmayer} {et~al.}(1998){Strohmayer}, {Zhang}, {Swank}, {White},
  \& {Lapidus}}]{1998ApJ...498L.135S}
{Strohmayer}, T.~E., {Zhang}, W., {Swank}, J.~H., {White}, N.~E., \& {Lapidus},
  I. 1998, \apjl, 498, L135

\bibitem[{{Weinberg} {et~al.}(2001){Weinberg}, {Miller}, \&
  {Lamb}}]{2001ApJ...546.1098W}
{Weinberg}, N., {Miller}, M.~C., \& {Lamb}, D.~Q. 2001, \apj, 546, 1098

\bibitem[{{Wijnands} {et~al.}(2001){Wijnands}, {Strohmayer}, \&
  {Franco}}]{2001ApJ...549L..71W}
{Wijnands}, R., {Strohmayer}, T., \& {Franco}, L.~. 2001, \apjl, 549, L71

\bibitem[{{Woosley} \& {Taam}(1976)}]{1976Natur.263..101W}
{Woosley}, S.~E. \& {Taam}, R.~E. 1976, \nat, 263, 101

\bibitem[{{Zingale} {et~al.}(2001){Zingale}, {Timmes}, {Fryxell}, {Lamb},
  {Olson}, {Calder}, {Dursi}, {Ricker}, {Rosner}, {MacNeice}, \&
  {Tufo}}]{2001ApJS..133..195Z}
{Zingale}, M., {Timmes}, F.~X., {Fryxell}, B., {Lamb}, D.~Q., {Olson}, K.,
  {Calder}, A.~C., {Dursi}, L.~J., {Ricker}, P., {Rosner}, R., {MacNeice}, P.,
  \& {Tufo}, H.~M. 2001, \apjs, 133, 195

\end{thebibliography}

\end{document}